\documentclass[journal]{IEEEtran}
\usepackage[T1]{fontenc}
\usepackage[utf8]{inputenc}
\usepackage{amsmath,amssymb}
\usepackage{booktabs}
\usepackage{multirow}
\usepackage{graphicx}
\usepackage{xcolor}
\usepackage{cite}
\usepackage{url}
\usepackage{microtype}
\usepackage{tikz}
\usetikzlibrary{positioning,arrows.meta,decorations.pathreplacing,calc,fit,shapes.geometric}
\usepackage{pgfplots}
\pgfplotsset{compat=1.17}
\usepgfplotslibrary{groupplots}
\definecolor{cmkc}{HTML}{2A78D6}\definecolor{ckmc}{HTML}{D9622B}\definecolor{cjf}{HTML}{2F9E77}\definecolor{chg}{HTML}{8E5BD0}
\pgfplotsset{
  paper/.style={width=\linewidth, height=4.1cm, tick label style={font=\scriptsize}, label style={font=\footnotesize},
    title style={font=\footnotesize, yshift=-1.2ex}, grid=major, grid style={gray!22, line width=0.3pt}, axis line style={gray!70},
    tick style={gray!70}, legend style={font=\scriptsize, draw=none, fill=none, legend columns=-1, column sep=3pt}, every axis plot/.append style={line width=0.9pt, mark size=1.7pt}},
  smkc/.style={cmkc, mark=*}, smkcb/.style={cmkc, mark=o, densely dashed}, skmc/.style={ckmc, mark=square*}, sjf/.style={cjf, mark=triangle*}, shg/.style={chg, mark=diamond*}}

\newcommand{\mkc}{\textsc{mkc}}
\newcommand{\kmc}{KMC\,3}
\newcommand{\jf}{Jellyfish\,2}
\newcommand{\hg}{Hackgap}
\newcommand{\kmer}{$k$-mer}
\newcommand{\kmers}{$k$-mers}
\newcommand{\rc}[1]{\overline{#1}}
\newcommand{\x}{$\times$}

\newcommand{\BatchHgSpreadMax}{8.5}
\newcommand{\BatchHgSpreadMin}{1.6}
\newcommand{\BitsHuman}{42.5}
\newcommand{\BitsHumanPacked}{37.4}
\newcommand{\BitsHumanRecords}{42.5}
\newcommand{\BucketLoad}{94}
\newcommand{\BuildSpeedHgMax}{42.4}
\newcommand{\BuildSpeedHgMin}{5.6}

\newcommand{\BuildSpeedKmcMax}{8.0}
\newcommand{\BuildSpeedKmcMin}{1.6}
\newcommand{\DenseMaxK}{15}
\newcommand{\ExactKmers}{38.8}
\newcommand{\HumanUnpacked}{Hs38}
\newcommand{\LargestOcc}{91}

\newcommand{\LikeHugeHgNaturalMax}{1.33}
\newcommand{\LikeHugeHgNaturalMin}{0.90}
\newcommand{\LikeHugeHgRandomMTMax}{3.6}
\newcommand{\LikeHugeHgRandomMTMin}{1.7}
\newcommand{\LikeHugeHgRandomMax}{3.3}
\newcommand{\LikeHugeHgRandomMin}{1.8}
\newcommand{\LikeHugeHgSpreadMTMax}{2.49}
\newcommand{\LikeHugeHgSpreadMTMin}{0.94}
\newcommand{\LikeHugeHgSpreadMax}{1.89}
\newcommand{\LikeHugeHgSpreadMin}{0.90}

\newcommand{\LikeHugeJfRandomMax}{11.3}
\newcommand{\LikeHugeJfRandomMin}{6.4}

\newcommand{\LikeHugeKmcRandomMax}{14.0}
\newcommand{\LikeHugeKmcRandomMin}{5.2}

\newcommand{\LikeSmallHgRandomMax}{3.3}
\newcommand{\LikeSmallHgRandomMin}{1.8}

\newcommand{\LikeSmallHgSpreadMax}{1.93}
\newcommand{\LikeSmallHgSpreadMin}{1.00}

\newcommand{\LikeSmallJfRandomMax}{12.1}
\newcommand{\LikeSmallJfRandomMin}{6.1}

\newcommand{\LikeSmallKmcRandomMax}{11.7}
\newcommand{\LikeSmallKmcRandomMin}{5.0}

\newcommand{\MaxStash}{813}
\newcommand{\NumDatasets}{ten}

\newcommand{\QbitsHuman}{34}

\newcommand{\SizeRatioHgMax}{1.7}
\newcommand{\SizeRatioHgMin}{1.2}
\newcommand{\SizeRatioJfMax}{2.6}
\newcommand{\SizeRatioJfMin}{1.5}
\newcommand{\SizeRatioKmcMax}{2.1}
\newcommand{\SizeRatioKmcMin}{1.3}
\newcommand{\SlotsHuman}{13}

\newcommand{\TbitsHuman}{4}

\newcommand{\VerifiedTotal}{28}
\newcommand{\WbitsHuman}{39}

\begin{document}

\title{MKC: Exact $k$-mer Counting into a Compact Cache-Line Cuckoo Index}

\author{Kamila~Szewczyk%
\thanks{K. Szewczyk is with the Algorithmic Bioinformatics group, Department of Computer Science, Saarland University, Saarbr\"ucken, Germany, and with the Department of Mathematics, Saarland University, Saarbr\"ucken, Germany (e-mail: szewczyk@cs.uni-saarland.de; ORCID: 0000-0002-4002-9866).}%
\thanks{This work has been submitted to the IEEE for possible publication. Copyright may be transferred without notice, after which this version may no longer be accessible.}}

\markboth{arXiv}{Szewczyk: MKC: Exact $k$-mer Counting into a Compact Cache-Line Cuckoo Index}

\maketitle

\begin{abstract}
Querying the counts of individual $k$-mers is an essential step of assembly evaluation and metagenomic classification. Disk-based counters such as KMC\,3 count in bounded memory but write sorted lists that are slow to query, whereas hash-based counters such as Hackgap hold large in-memory tables for all distinct $k$-mers. In this paper, we propose an exact canonical $k$-mer counter called \mkc{} for $k\le 32$, which builds a static hash table of counts in the manner of a disk-based counter. \mkc{} distributes the $k$-mers into bins by minimizer and counts each bin by sorting the hashes of its $k$-mers in cache. Subsequently, it assembles a two-choice cuckoo table of 64-byte buckets, in which each record stores a quotient of an invertible hash and a count field of variable width, and buckets of mostly unique $k$-mers store their quotients via Elias--Fano coding. We apply \mkc{} on ten genomes and read sets. The results show that the index of a human genome takes \BitsHumanPacked--\BitsHuman{} bits per distinct 31-mer, counts included, and that most lookups read one cache line. With 16 threads, \mkc{} builds its index \BuildSpeedKmcMin--\BuildSpeedKmcMax{} times faster than KMC\,3 and \BuildSpeedHgMin--\BuildSpeedHgMax{} times faster than Hackgap, indexes a human genome in 8.9\,s with a peak resident set size of 1.0\,GiB, and writes the smallest index of the compared counters. At equal page size, it answers random queries \LikeSmallKmcRandomMin--\LikeSmallKmcRandomMax{} times faster than KMC\,3 and \LikeSmallHgRandomMin--\LikeSmallHgRandomMax{} times faster than Hackgap. For $k$-mers in sequence order, its batch interface is \BatchHgSpreadMin--\BatchHgSpreadMax{} times faster than Hackgap.
\end{abstract}


\section{Introduction}\label{sec:intro}
\IEEEPARstart{C}{ounting} the occurrences of all substrings of length $k$ (\kmers) in a collection of sequences is an essential task in sequence analysis. \kmer{} counts are used to estimate genome size, heterozygosity and ploidy~\cite{genomescope2}, to select parameters for genome assembly~\cite{kmergenie}, and to evaluate the quality and completeness of an assembly without a reference~\cite{merqury}. Many applications do not only need the distribution of the counts but also query the counts of individual \kmers. For example, Merqury~\cite{merqury} looks up every \kmer{} of an assembly in the \kmer{} counts of the reads, classifiers and decontamination filters~\cite{kraken2,xengsort} look up one \kmer{} per read position, and error correctors and variant genotypers look up the counts of specific \kmers. In such applications, the output of a \kmer{} counter is used as a static dictionary. Therefore, its size, the number of cache misses per lookup, the cost of a lookup of an absent \kmer, and whether several processes can share one copy of it in memory are as important as the speed of counting.

Existing counters address only one side of this problem. In-memory counters build a hash table that can be queried, but the table must hold all distinct \kmers{} during counting and every occurrence causes a random access to it. Disk-based counters count in bounded memory by partitioning the input, but they write sorted lists, on which a lookup costs several cache misses and unpredictable branches. To our knowledge, no exact counter builds, in bounded memory, an index that answers most lookups of stored and absent \kmers{} with a single cache line.

In this work, we propose \mkc, an exact counter of canonical \kmers{} for $1\le k\le 32$ that builds a static hash table of counts in the manner of a disk-based counter. \mkc{} splits the input into runs of \kmers{} with the same minimizer, counts each bin by sorting the final table hashes of its \kmers{} in cache, writes the sorted hashes to a spool of Rice-coded slices, and then assembles the table from the spool group by group, each group in cache. Because the whole key set of a group is known before any record is placed, \mkc{} can place frequent \kmers{} in their first-choice bucket, set exact overflow filters, fix the load at \BucketLoad\% in advance, and choose the width of the count field from the count distribution. The resulting index is a two-choice cuckoo table of 64-byte buckets, in which each record stores only a quotient of an invertible hash, one choice bit and a count field, and in which buckets of mostly unique \kmers{} store up to 15 quotients as an Elias--Fano sequence.

The contributions of this paper are as follows. First, we introduce a cache-line cuckoo index for \kmer{} counts whose records store quotients, choice bits and count fields of variable width, with overflow filters that stop most misses after one cache line (Section~\ref{sec:index}). Second, we present an offline construction that makes random memory accesses only within the cache, limits its counting workspace by a budget, and writes a byte-identical index regardless of thread scheduling (Section~\ref{sec:build}). Third, we compare \mkc{} with \kmc, \jf, \hg{} and weighted SSHash on \NumDatasets{} genomes and read sets with up to \LargestOcc{} billion \kmer{} occurrences, verify every index exactly, and analyze each design decision in an ablation study (Section~\ref{sec:eval}). Finally, we release the implementation together with a harness that reproduces every measurement.

\section{Related Work}\label{sec:related}
Most early counters work in memory. Jellyfish~\cite{jellyfish} counts \kmers{} in a lock-free open-addressing hash table that stores only the part of each key that is not implied by its slot under an invertible hash. BFCounter~\cite{bfcounter} and Turtle~\cite{turtle} remove most singleton \kmers{} with an approximate filter before counting, Squeakr~\cite{squeakr} counts in a counting quotient filter~\cite{cqf}, and CHTKC~\cite{chtkc} in a lock-free chained hash table. The memory of these methods grows with the number of distinct \kmers, and their speed is limited by cache misses.

Disk-based counters partition the input before counting. DSK~\cite{dsk} and the first version of KMC~\cite{kmc1} distribute the \kmers{} into files by hash or by prefix and count one file at a time. MSPKmerCounter~\cite{msp}, KMC\,2~\cite{kmc2} and Gerbil~\cite{gerbil} distribute super-\kmers, i.e., runs of consecutive \kmers{} with the same minimizer, which reduces the temporary data several-fold. \kmc~\cite{kmc3} adds a faster radix sort and faster input handling, and independent benchmarks rank it first on large inputs~\cite{manekar}. Other recent counters include KCOSS~\cite{kcoss} for assembled genomes, BIKE~\cite{bike}, whose memory does not depend on the alphabet and whose parallelism is deterministic, and counters for distributed memory~\cite{kmerind,pan_sc18,hysortk}. The output of \kmc, like that of Jellyfish, is a sorted list of \kmers, which is queried through a prefix table and a binary search.

\hg~\cite{hackgap} counts \kmers, optionally gapped, directly into a three-way bucketed cuckoo hash table~\cite{zentgraf_cost} with quotiented keys, and xengsort~\cite{xengsort} shows that such a table can be queried quickly. \hg{} inserts the \kmers{} online, so the capacity of the table must be specified in advance, and the whole table is held in memory and accessed at random during counting. Both of its techniques are well established. Cuckoo hashing~\cite{cuckoo} restricts each key to two positions and, with buckets of several slots, reaches very high loads~\cite{dietzfelbinger,fotakis}. Quotienting, introduced by Knuth and made practical by Cleary~\cite{cleary}, stores only the part of an invertible hash of a key that its position does not imply, and it is used by quotient filters~\cite{quotientfilter,cqf} and, together with cuckoo hashing, by cuckoo filters~\cite{cuckoofilter}, which store short fingerprints and answer approximately.

Compressed \kmer{} dictionaries exploit the overlaps between the \kmers{} of a genome. They store a set of \kmers{} as strings that spell each \kmer{} once and locate a \kmer{} through a minimal perfect hash~\cite{bbhash} of its minimizer. SSHash~\cite{sshash} is the leading example. Its weighted variant~\cite{sshashw} also stores exact counts, run-length compressed along the strings, in about nine bits per \kmer{} for a human genome, and a recent redesign~\cite{sshashopt} reduces the cache misses per lookup. SSHash is built from the output of a counter by a separate unitig construction step, and a lookup costs several cache misses and a comparison with the stored strings.

In summary, in-memory counters are queryable but need memory for all distinct \kmers, disk-based counters bound their memory but produce lists that are slow to query, and compressed dictionaries are small but slow to build. \mkc{} combines the construction of a disk-based counter with a hash index whose lookups usually read one cache line.

\section{The Proposed MKC Method}\label{sec:methods}
Section~\ref{sec:prelim} defines the problem, Section~\ref{sec:index} describes the structure of the index, and Section~\ref{sec:build} describes its construction (Fig.~\ref{fig:pipeline}).

\subsection{Background and Notation}\label{sec:prelim}
We identify a DNA \kmer{} $x$ with the $2k$-bit integer obtained from the encoding $\mathtt{A}{=}0,\mathtt{C}{=}1,\mathtt{G}{=}2,\mathtt{T}{=}3$, with the first base most significant, and we write $\rc{x}$ for the reverse complement of $x$. Since a molecule can be sequenced from either strand, counters identify $x$ with $\rc{x}$ and count \emph{canonical} \kmers, usually represented by $\min(x,\rc{x})$. Let $N$ be the number of \kmer{} occurrences of an input, i.e., of windows of $k$ consecutive bases from $\{\mathtt{A},\mathtt{C},\mathtt{G},\mathtt{T}\}$, and let $n$ be the number of distinct canonical \kmers. Our goal is to build a structure that returns the exact number of occurrences of any \kmer, and zero for a \kmer{} that does not occur. We consider $k\le 32$, so that a \kmer{} fits in a machine word. This range includes the values used by the applications above, which are typically between 21 and 31.

The \emph{minimizer} of a \kmer{} is its smallest substring of length $m$ under a given order~\cite{roberts,winnowing}. Since consecutive \kmers{} often share their minimizer, a read can be decomposed into few maximal runs of \kmers{} with the same minimizer, called super-\kmers, and a super-\kmer{} takes far less space than its \kmers{} stored individually~\cite{msp,kmc2}.

\subsection{Index Structure}\label{sec:index}
An \mkc{} index is one file that consists of a 128-byte header, a table, and an appendix of exceptional counts. The file is immutable, position-independent and little-endian. We expect programs to use it by mapping it into memory through a dependency-free, C99 header file. The index has two layouts: the \emph{hash} layout described in this subsection and a \emph{dense} array for small $k$ (Section~\ref{sec:dense}). \mkc{} writes whichever of the two is smaller.

\subsubsection{Canonical Codes}\label{sec:codes}
For even $k$, \mkc{} uses the $2k$-bit code $c(x)=\min(x,\rc{x})$. For odd $k$, no \kmer{} equals its reverse complement, so there are exactly $4^k/2$ canonical \kmers, and we use this property to save one bit. Specifically, complementation flips both bits of the middle base, so the high bit of the middle base is clear in exactly one of $x$ and $\rc{x}$. \mkc{} chooses this orientation and deletes the bit, which yields a bijection between canonical \kmers{} and $(2k{-}1)$-bit integers. In addition to saving one bit per record, this rule decides the orientation by a single bit test instead of a comparison of two words. We write $b$ for the code width, i.e., $b=2k-1$ for odd $k$ and $b=2k$ for even $k$.

\subsubsection{Invertible Hashing and Quotienting}
\mkc{} scrambles the codes with a bijection $\pi_b$ on $b$-bit integers. $\pi_b$ consists of an xor with a constant, two rounds of an xor-shift followed by a multiplication with an odd constant modulo $2^b$, and a final xor-shift, each of which is invertible. Therefore, the hash $h=\pi_b(c(x))$ identifies the \kmer{} exactly, no two \kmers{} collide, and $x$ can be recovered from $h$, which \mkc{} uses to enumerate the index.

Note that $\pi_b$ is one fixed function and not a random member of a hash family, so key sets that it maps into any chosen part of its range are easy to construct. We therefore assume that the hashes of the \kmers{} of real sequences spread over the range like random values, which affects only efficiency and which we test in Section~\ref{sec:anatomy} and Section~\ref{sec:robust}. All structures remain exact on any key set, because keys that do not fit where the assumption would place them are stored in exception lists.

The table is divided into $2^g$ \emph{groups} of $C$ buckets each, where $C\le 4096$. The top $g$ bits of $h$ select the group, and the remaining $s=b-g$ bits form the suffix $v_0$. A key has two candidate buckets in its group, bucket $v_0 \bmod C$ (\emph{choice~0}) and bucket $v_1 \bmod C$ (\emph{choice~1}), where $v_1=\pi_s(v_0)$ is a bijection of the same kind on $s$ bits. A key placed under choice $i$ is represented by the quotient $\lfloor v_i/C\rfloor$ and the bit $i$, which together with the bucket number determine $v_i$, and hence $v_0$, $h$ and the \kmer. A record is thus
\[
 \underbrace{\lfloor v_i/C\rfloor}_{q\ \text{bits}}\;\Big|\;\underbrace{\,i\,}_{1\ \text{bit}}\;\Big|\;\underbrace{\text{tag}}_{t\ \text{bits}},\qquad q=\left\lceil \log_2 (2^s/C) \right\rceil ,
\]
where the tag encodes the count (Section~\ref{sec:counts}). Since $C$ need not be a power of two, \mkc{} can match the size of the table to the key set within a fraction of a percent, and it divides by $C$ through a multiplication with a precomputed reciprocal~\cite{barrett}.

Groups are used during construction (Section~\ref{sec:assemble}). A group occupies at most $64C\le 256$\,KiB, so all displacements during assembly stay within the second-level cache, and different threads can build different groups independently.

\subsubsection{Cache-Line Buckets}
\begin{figure}[t]
\centering
\begin{tikzpicture}[font=\footnotesize, x=0.068\columnwidth, y=0.6cm]
  % bucket of 12 slots
  \foreach \i in {0,...,11} {
    \draw[fill=gray!10] (\i*1.15,0) rectangle ++(1.15,1);
    \node at (\i*1.15+0.575,0.5) {\scriptsize \i};
  }
  \draw[fill=gray!45] (13.8,0) rectangle ++(0.5,1);
  \node[anchor=west] at (-0.1,1.45) {64-byte bucket: e.g.\ 12 slots of 42 bits};
  \node[anchor=east] at (14.4,1.45) {8 spare bits};
  \draw[-{Stealth[length=1.5mm]}] (14.05,-0.9) -- (14.05,-0.05);
  \node[anchor=east] at (14.4,-1.15) {overflow filter};
  % zoom
  \draw[densely dotted] (2.3,0) -- (0.0,-2.0);
  \draw[densely dotted] (3.45,0) -- (11.2,-2.0);
  \draw[fill=blue!12] (0,-3.0) rectangle ++(6.9,1);
  \draw[fill=orange!30] (6.9,-3.0) rectangle ++(0.8,1);
  \draw[fill=green!18] (7.7,-3.0) rectangle ++(3.5,1);
  \node at (3.45,-2.5) {quotient $\lfloor v_i/C\rfloor$};
  \node at (7.3,-2.5) {$i$};
  \node at (9.45,-2.5) {tag};
  \draw[decorate,decoration={brace,mirror,amplitude=3pt}] (0,-3.1) -- node[below=3pt] {$q$ bits} (6.9,-3.1);
  \draw[decorate,decoration={brace,mirror,amplitude=3pt}] (7.7,-3.1) -- node[below=3pt] {$t$ bits} (11.2,-3.1);
  \node[anchor=west,align=left] at (0,-4.75) {tag ($t\ge 7$): $0$ empty $\mid$ $1..2^{t-1}{-}1$ count $\mid$ $2^{t-1}..2^t{-}2$ index of an\\ exact count in the group's array $\mid$ $2^t{-}1$ count in the dictionary};
\end{tikzpicture}
\caption{Bucket and record layout of the hash index for 12 slots. For the human genome (\HumanUnpacked) at $k=31$, \mkc{} chooses \SlotsHuman{} slots of \WbitsHuman{} bits with $q=\QbitsHuman$ and $t=\TbitsHuman$. A lookup compares the low $q{+}1$ bits of all slots with the quotient and choice bit of the query at once.}
\label{fig:bucket}
\end{figure}
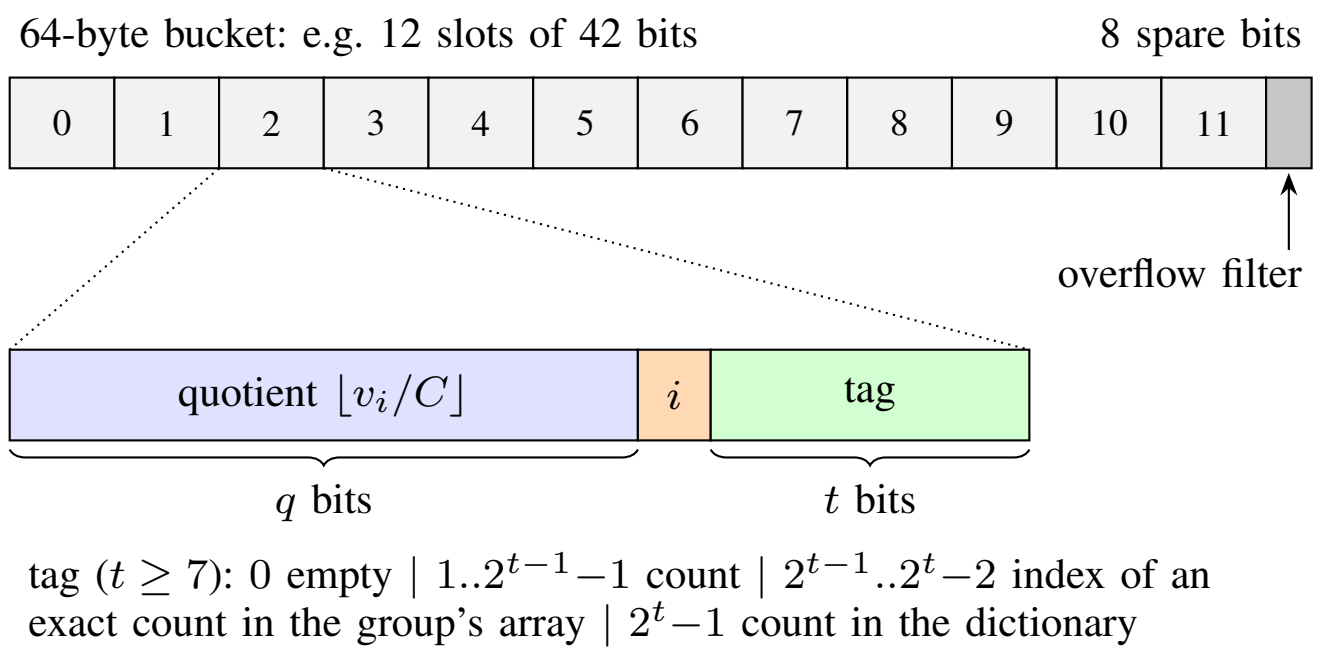

A bucket has 64 bytes and holds $S\in\{8,\dots,14\}$ records of $W=\lfloor 512/S\rfloor\in\{64,56,51,46,42,39,36\}$ bits, packed from bit~0 (Fig.~\ref{fig:bucket}). To choose the shape of the table, \mkc{} enumerates $S$, $g$ and the tag width $t$ for the given $n$ and count distribution, and it sets $C=\lceil n/(0.94\cdot 2^g S)\rceil$. If the quotient would not leave $t$ bits for the tag, \mkc{} increases $C$, and it discards shapes that would need $C>4096$. Among the remaining shapes, \mkc{} keeps the one for which the table and the exceptional counts that the tag cannot hold (Section~\ref{sec:counts}) are smallest together. Large key sets have short quotients. For example, for a human genome, in which few counts are large, \mkc{} chooses thirteen 39-bit slots with a four-bit tag, while small key sets and deep read sets get fewer and wider slots.

A table of $n$ keys takes $512/(0.94\,S)$ bits per key including small counts, i.e., $41.9$ bits for $S=13$. The information-theoretic minimum for a set of $n$ keys out of $2^b$ is about $\log_2(2^b/n)+1.44$ bits per key, and the key part of a record, i.e., the quotient and the choice bit divided by the load, is within about 6 bits of it. All remaining bits of the record hold the count. This bound holds for any dictionary that treats \kmers{} as unrelated keys and resolves each key in one cache line, but not for compressed dictionaries, which exploit the overlaps between \kmers{} (Section~\ref{sec:sshash}).

\subsubsection{Counts}\label{sec:counts}
Counts are exact 64-bit values, but nearly all of them are small. All bits of a record after the quotient and the choice bit form the tag, i.e., $t=W-q-1$, capped at 24. A tag of at least seven bits is split into two ranges. The values $1,\dots,2^{t-1}-1$ are the count itself, and the values $2^{t-1},\dots,2^t-2$ index an array of 64-bit counts appended to the index, relative to a per-group offset, so that up to $2^{t-1}-1$ large counts per group cost two loads and no search. Only if a group has more large counts does the tag take the value $2^t-1$, which refers to an open-addressing dictionary keyed by $h$. A shorter tag of two to six bits holds counts up to $2^t-2$, and larger counts go to the dictionary. \mkc{} chooses the tag width from the histogram of the counts. For example, only 0.16\% of the \kmers{} of a human genome occur more than 14 times, so a four-bit tag makes room for a thirteenth record per bucket at the cost of four million dictionary entries, whereas in a read set of 30\x{} coverage most occurrences belong to \kmers{} with counts above 14, and the tag remains wide.

\subsubsection{Overflow Filter and Stash}
In a two-choice table, a lookup that does not find its key in the first bucket must usually inspect the second bucket. To avoid most of these accesses, \mkc{} uses the bits that the slots of a bucket leave over ($512-SW$, between two and eight) as an \emph{overflow filter}. Every key whose choice-0 bucket is this bucket but which is stored elsewhere sets one of these bits, selected by a hash of its quotient. After a miss in the first bucket, a lookup tests the bit selected by the wanted quotient, and if the bit is clear, the key is absent. A single bit would be a flag that sends every miss in a bucket with displaced keys on to the second bucket, whereas with five or eight bits most of these misses stop, because such a bucket has displaced only two or three keys (Section~\ref{sec:ablation}). In the 64-bit shape, which has no spare bit, the last slot takes the role of the flag. Keys that fit in neither bucket (Section~\ref{sec:assemble}) are stored in a \emph{stash}, which shares the dictionary and is consulted only after two misses with the filter bit set. With the default settings, the stash held at most \MaxStash{} keys on real data in all our experiments.

\subsubsection{Packed Buckets}\label{sec:packed}
A bucket of $S$ records stores every quotient in full, although the set of its quotients carries about $\log_2 S!$ bits less information than their sequence, and if most \kmers{} occur once, most of the count field is unused. To remove both redundancies, \mkc{} can use \emph{packed} buckets, each still one self-contained cache line. Let $v=i\cdot 2^q+\lfloor v_i/C\rfloor$ be the value of a key in $u=q+1$ bits. A packed bucket stores the values of its keys as a sorted Elias--Fano sequence~\cite{elias_ef,fano} after a 6-bit header: the top four bits of each value in unary ($S+16$ bits) and the low $L=u-4$ bits at fixed offsets. A list of at most three entries follows, each holding the 4-bit position and the tag of a key whose count is not one, and the bits between the low parts and the list form an overflow filter of at least six bits. A packed bucket holds as many keys as fit, at most 15, against 13 records for the human genome.

\mkc{} sizes a packed table for the mean capacity of a bucket, which follows from the binomial distribution of the list entries, and uses packed buckets only if at most one list entry is expected in a full bucket and the table is at least 3\% smaller. For HsT2T, packed buckets reduce the index from \BitsHumanRecords{} to \BitsHumanPacked{} bits per \kmer, whereas the read sets, CoV, \emph{P.~taeda} and Hs38 have many repeated \kmers{} and keep the record layout. Placement works on an array of 16 words per bucket in groups of at most 2048 buckets, which stays in the second-level cache, and each bucket is then sorted with a 16-input sorting network~\cite{batcher} and encoded. A lookup compares all low parts with the wanted one at once and accepts a matching slot $j$ if its top bits, i.e., the position of the $j$-th set bit of the unary part minus $j$, computed with one \texttt{pdep} instruction, equal the wanted ones. A SWAR comparison then tests whether the key has a list entry, and for the many keys without one a well-predicted branch returns a count of one.

\subsubsection{Lookup}\label{sec:lookup}
A point query computes $c(x)$, $h$, the group and $v_0$, loads one bucket, and compares all slots with the wanted $q{+}1$ low bits at once. Three 256-bit vectors cover twelve slots, each lane gathering one slot with a byte shuffle and a variable shift, and further slots are compared as words. If a slot matches and its tag is an inline count, this count is the answer, and if none matches and the filter bit is clear, the answer is zero. The second bucket, large counts and the stash are handled out of line. The reader instantiates this path for every pair of $k$ and record width (224 kernels) and, for packed buckets, for every $L$ from 20 to 40, and selects one kernel when the index is opened, so that all offsets and shifts are immediate constants. The resulting straight-line code of about 150 instructions lets an out-of-order core overlap the cache misses of consecutive queries. A portable scalar path gives identical results on other CPUs, and a batch interface processes blocks of 64 keys in a software pipeline that requests the buckets of the next block, and the second cache line of the lookups that need one, before it completes the current block.

File mappings use 4\,KiB pages on common systems, and random lookups in a table of several gigabytes then also miss in address translation. Therefore, if an index has at least 64\,MiB and at most half of the available memory, the reader by default copies it into anonymous memory backed by 2\,MiB pages, while \texttt{mkc\_open\_mapped} maps the file, which is available at once and shared by all processes (Section~\ref{sec:query}).

\subsubsection{Dense Layout}\label{sec:dense}
If $2^b$ is small compared with $n$, a direct array of counts indexed by $c(x)$ is smaller and faster than any hash table. The dense layout stores $2^b$ counters of 1, 2, 4 or 8 bytes, whichever minimizes the file, and stores larger counts in the dictionary, so that a query is one load. \mkc{} compares the sizes of both layouts and writes the smaller one, which makes the dense layout the choice for $k\le\DenseMaxK$ on the human genome.

\subsection{Construction}\label{sec:build}
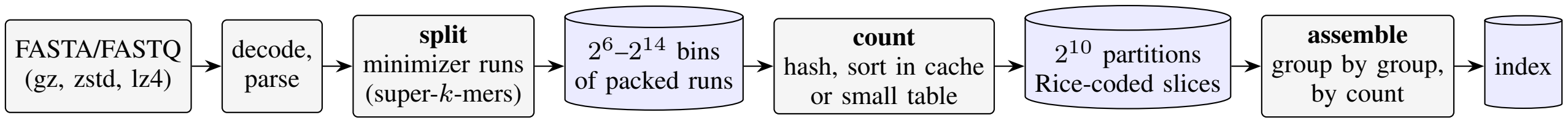
\begin{figure*}[t]
\centering
\begin{tikzpicture}[font=\footnotesize, node distance=3.2mm,
  box/.style={draw, rounded corners=1.5pt, align=center, minimum height=10mm, inner sep=3pt, fill=gray!8},
  disk/.style={draw, align=center, minimum height=10mm, inner sep=3pt, fill=blue!8, cylinder, shape border rotate=90, aspect=0.12},
  ar/.style={-{Stealth[length=1.8mm]}}]
  \node[box] (in) {FASTA/FASTQ\\(gz, zstd, lz4)};
  \node[box, right=of in] (parse) {decode,\\parse};
  \node[box, right=of parse] (split) {\textbf{split}\\minimizer runs\\(super-\kmers)};
  \node[disk, right=of split] (bins) {$2^{6}$--$2^{14}$ bins\\of packed runs};
  \node[box, right=of bins] (count) {\textbf{count}\\hash, sort in cache\\or small table};
  \node[disk, right=of count] (spool) {$2^{10}$ partitions\\Rice-coded slices};
  \node[box, right=of spool] (asm) {\textbf{assemble}\\group by group,\\by count};
  \node[disk, right=of asm] (idx) {index};
  \draw[ar] (in) -- (parse); \draw[ar] (parse) -- (split); \draw[ar] (split) -- (bins); \draw[ar] (bins) -- (count);
  \draw[ar] (count) -- (spool); \draw[ar] (spool) -- (asm); \draw[ar] (asm) -- (idx);
\end{tikzpicture}
\caption{Construction pipeline of \mkc{} for $k\ge 14$. All stages read and write sequentially. Random accesses occur only within regions that fit in the cache (a counting bucket, a table group) and in the dictionary of exceptional counts, which is built last.}
\label{fig:pipeline}
\end{figure*}

For $k\le 13$, the code space has at most $2^{25}$ elements, and \mkc{} counts directly into an array, staging the increments per 64\,KiB region so that the updates stay in cache. For larger $k$, construction proceeds in three stages, split, count and assembly (Fig.~\ref{fig:pipeline}), each of which accesses data only sequentially or within regions of cache size.

\subsubsection{Input}
One reader thread per input file decodes the file into a ring of 8\,MiB blocks ahead of the parser. \mkc{} recognizes the format from the first bytes and inflates gzip with ISA-L. BGZF members~\cite{tabix}, zstd frames that declare their content size, and lz4 frames of independent blocks are decoded by all threads in parallel. FASTQ and FASTA are also parsed in parallel, in slices of a block that resynchronize on record boundaries and must tile the block exactly, and in any unusual case \mkc{} falls back to a serial parser with identical semantics. Bases are translated to two-bit codes with AVX2, and any other byte (e.g., \texttt{N}) breaks the sequence.

\subsubsection{Split}\label{sec:split}
In the first stage, scanner threads cut the sequence into runs of consecutive \kmers{} whose minimizers map to the same bin. \mkc{} uses canonical $m$-mers with $m=\min(12,\max(9,k-8))$, ordered by a \emph{double decycling set}~\cite{mykkeltveit,decycling}. Let $w(x)=\sum_q x_q\sin(2\pi q/m)$ for an $m$-mer $x$, and let $r(x)$ be its rotation by one place. The $m$-mers with $w(x)>0\ge w(r(x))$ rank first, those with $w(x)<0\le w(r(x))$ second, all others last, and a hash orders each tier; $w$ is computed in fixed point, and $|w|\le 2^{-8}$ counts as zero. The first set is the decycling set of Mykkeltveit~\cite{mykkeltveit} without the $m$-mers that it adds for rotation classes of zero weight, so cycles of the de~Bruijn graph of order $m$ through such classes can avoid it, e.g., the self-loop of $A^m$ or, for even $m$, the cycle of alternating A and C. Moreover, \mkc{} ranks canonical $m$-mers, and the canonical form of an $m$-mer of a sequence need not lie in the set, so this property motivates the order but does not carry over to it. Any order gives a correct partition, and we measure its effect instead: on our inputs, it produces 6--8\% fewer runs and 5--8\% smaller bins than a hash order, and, unlike the lexicographic order~\cite{kmc2}, it does not produce skewed bins on low-complexity sequence. \mkc{} computes the window minima for tiles of 2048 positions without data-dependent branches, eight positions at a time with AVX2 by $\lfloor\log_2(k-m+1)\rfloor$ doubling passes, and otherwise from prefix and suffix minima over blocks~\cite{vanherk,gilwerman}; both give the same minimizers.

A run is written bit-packed as the number of further \kmers{} (6 bits), its first \kmer{} ($2k$ bits) and up to 32 further bases, i.e., 68 bits for a single 31-mer and 2 bits for each additional \kmer, into per-thread blocks of the bin selected by a hash of the minimizer. Since the minimizer is canonical, both orientations of a \kmer{} reach the same bin. The number of bins is a power of two between $2^{6}$ and $2^{14}$, at least $2^{10}$ and large enough that an average bin fits the workspace share of a thread, except for small inputs, which get about 256k occurrences per bin.

\subsubsection{Count}\label{sec:count}
In the second stage, worker threads take the bins in decreasing order of size. The key idea of this stage is to sort the \emph{final table hashes} $h$ of the \kmers{} instead of the \kmers{} themselves. Since $\pi_b$ is a bijection, a sorted array of 8-byte hashes is the counted bin, already in table order, and on real sequences its values are spread evenly over the hash range regardless of the base composition. \mkc{} produces this array along one of three paths.

\emph{Table path.} For deep coverage or collections of near-identical genomes, a bin has few distinct \kmers{} per occurrence, and \mkc{} counts it in a small open-addressing table of at most $2^{18}$ entries, so that memory and sorting scale with $n$ instead of $N$. It abandons the attempt as soon as the distinct \kmers{} exceed a quarter of the occurrences seen or the table would exceed the cache, and after 32 consecutive failures it tries only every 64th bin.

\emph{Sort path.} Otherwise, \mkc{} distributes the hashes by their top 11 bits into 2048 ranges. A textbook radix pass decodes the bin twice, once for the histogram and once to scatter, but for evenly spread hashes the histogram is known in advance up to sampling noise. \mkc{} therefore gives each range a region for its expected $\mu=n'/2048$ keys plus $8\sqrt{\mu}$ and decodes the bin \emph{once}, routing each hash through one cache line of staging per range that is flushed with non-temporal stores~\cite{wassenberg,balkesen}. A range inflated by a highly repeated \kmer{} spills its surplus to a shared list, which holds 5.8\% of the keys of a human genome, and only if this list fills does the bin fall back to two passes. Each range is then sorted in cache with an MSD radix sort whose digit width adapts to the range size, and two adjacent ranges are encoded into one spool partition while they are still in cache. In a deep read set, the sorter detects on the first range of a bin that the ranges hold few distinct keys, and from then on it counts each range in a small table and sorts only the distinct keys.

\emph{Passes.} A repeat whose minimizer is shared by many copies, or the low density of the decycling order, makes some bins several times larger than the mean. \mkc{} counts a bin of more than 1.5 times the mean, or of more than the workspace share of a thread, in passes over contiguous ranges of spool partitions, each of which decodes the whole bin and keeps only the hashes of its range. A partition lies within one pass, so the index does not change, while the memory for a bin is bounded by the size of a pass. Only a single partition larger than the workspace share, such as one \kmer{} repeated tens of millions of times, is sorted in chunks that are merged.

\subsubsection{Spool}\label{sec:spool}
\mkc{} collapses the sorted hashes of a bin into (hash, count) pairs and appends them to a spool of $2^{10}$ partition files, given by the top ten bits of the hash. A bin contributes one \emph{slice} to each partition, which holds the low $b-10$ bits of its hashes in increasing order: the first hash raw, each subsequent one as the Golomb--Rice code~\cite{golomb,rice} of the gap to its predecessor, with the parameter chosen from the density of the slice. A count of one costs a single bit, and larger counts are Elias-$\gamma$ codes~\cite{elias}. For evenly spread hashes, the gaps are close to geometric, and the code is within a fraction of a bit of their entropy. The spool replaces the in-memory arrays of (\kmer, count) pairs that a sorting counter holds or merges, which is why the resident memory stays far below the size of a large table.

\subsubsection{Assembly}\label{sec:assemble}
In the third stage, each assembly worker takes a unit of whole groups (one partition, or several if groups are larger), maps its partition files, and decodes them group by group. Since every slice is sorted, the keys of the next group form a prefix of the remainder of every slice, and these prefixes together yield the key set of the group without sorting or merging. The worker then places the at most $4096\cdot S$ keys of the group into its region of at most 256\,KiB in four steps.
\begin{enumerate}
\item The keys are ordered by decreasing $\lfloor\log_2 \mathrm{count}\rfloor$ and offered their choice-0 bucket in this order. Keys of count one, which are the large majority, form the last tier and are kept in a list of their own, and only the other keys pass through a stable counting sort. Frequent \kmers{} therefore occupy their first-choice buckets, which matters because queries derived from sequence hit \kmers{} in proportion to their counts.
\item A key whose choice-0 bucket is full takes a free slot of its choice-1 bucket if one exists.
\item Otherwise, \mkc{} tries to free a slot in either bucket by moving a resident key that sits in its first-choice bucket to its second-choice bucket, if that bucket has a free slot. Residents are examined from the last slot, i.e., the least frequent first.
\item As a last resort, \mkc{} attempts a bounded random-walk cuckoo insertion of 256 steps, and the key that is left over if it fails goes to the stash.
\end{enumerate}
Finally, \mkc{} sets the overflow filters of the group. Because the whole key set of a group is known before placement and the load is fixed in advance, steps 3 and 4 are rare, and the outcome does not depend on scheduling. The index is byte-for-byte identical across runs and across thread counts that lead to the same number of bins, provided that the lists of exceptional counts fit in the workspace, which all default builds in our experiments did. Two writer threads write the finished units at their final offsets with direct I/O, so that the table is written exactly once and no dirty pages compete with the spool for memory.

\subsubsection{Resources and Integrity}\label{sec:resources}
To limit the memory of construction, \mkc{} uses a workspace budget $M$, 6400\,MiB by default. With $T$ threads, each thread counts bins of up to $M/(16T)$ occurrences in an array of 8-byte words, and larger bins are counted in passes, so a thread holds at most about one mean bin at a time. Parsing and splitting take 16\,KiB per bin and about 4\,KiB per bin and scanner thread, and during assembly each thread holds one unit, about $1/1024$ of the index. Therefore, the resident memory grows slowly with $n$ and is a small fraction of the size of the table for large inputs. $M$ is a budget for the counting workspace and not a bound on the resident set size (Section~\ref{sec:memory}), and the bins and the spool are files, whose pages in the page cache the resident set size does not include. Every stage makes a bounded number of passes over its data, the radix sorts take linear time, and placement spends a bounded number of steps on a key before it uses the stash, so the running time is linear in $N$ on any input, and the spread of the hashes affects only the constant factor.

If more than 4\,096 keys and more than one key in 64 have been stashed, \mkc{} abandons the attempt and builds again with the next of four hash functions $\pi_b(c\oplus\sigma)$, recording the seed $\sigma$ in the header (Section~\ref{sec:robust}). To detect errors, the counting stage accumulates the number of distinct keys, the total count and a checksum over all (hash, count) pairs, and assembly recomputes all three from the records that it decodes, and \mkc{} refuses to publish the index if any value differs. The output is written under a temporary name and renamed after completion.

\section{Results}\label{sec:eval}
\subsection{Experimental Setup}\label{sec:setup}
All experiments run on one workstation with an AMD Ryzen~9 5950X (16 cores, 32 hardware threads, 64\,MiB L3 cache), 128\,GB of DDR4 memory and a 4\,TB WD Black SN850X NVMe SSD with ext4, which holds the inputs, the temporary files and the outputs, under Ubuntu~24.04 (Linux~6.8) with transparent huge pages in \texttt{madvise} mode. Unless stated otherwise, each tool runs with 16 threads, pinned to one hardware thread of each core.

We compare \mkc{} (commit \texttt{f46b4d4}, C99, GCC~13.3, \texttt{-O3}) with the tools that produce a queryable table of all \kmers{} from sequence, i.e., \kmc{} (3.2.4, release binary)~\cite{kmc3}, \jf{} (2.3.1, \texttt{-O3 -march=native})~\cite{jellyfish} and \hg~1.0.1 (Python~3.12, Numba~0.68)~\cite{hackgap}, and for index size and lookups also with weighted SSHash~\cite{sshashw,sshashopt}. Unless $k$ is varied, all tools count canonical 31-mers, keep every \kmer{} including singletons, and use counters wide enough for the largest count. \mkc{} runs with its defaults ($M=6400$\,MiB). We run \kmc{} with \texttt{-m4} and \texttt{-m16} and report the faster setting, give Jellyfish an initial hash size of $n/0.8$, so that its table is never resized, and give \hg{} the exact $n$, 13 subtables and its recommended thread split, where $n$ and the maximum count are taken from an \mkc{} index.

All tools were measured in a single session on the same machine. Every build is repeated three times, in rotating tool order, and the tables report the median and half the range. Before each build, we read the inputs once, flush dirty pages, and let the machine idle for 20\,s (60\,s for inputs above 2\,GB), and the harness records the share of input pages resident at the start and the bytes read from and written to the device. We measure the time until the index is durable, i.e., until the process exits and a subsequent \texttt{sync} completes, because \mkc{} writes with direct I/O, whereas the other tools leave dirty pages behind. Peak memory is the maximum resident set size of the process, which does not include the page cache, through which the temporary files of \mkc{} and \kmc{} pass. Table~\ref{tab:data} lists the data sets, and the supplementary material contains the harness, the download and setup scripts, and the raw results.

\begin{table}[t]
\caption{Data sets. $N$: \kmer{} occurrences, $n$: distinct canonical \kmers{} ($k=31$), both in billions.}
\label{tab:data}
\centering\footnotesize\setlength{\tabcolsep}{2pt}
\begin{tabular}{@{}llllrrr@{}}
\toprule
 & Data set & Accession & GB & $N$ & $n$ & $N/n$\\
\midrule
Ec & \emph{E. coli} & GCF\_000008865.2 & 0.006 & 0.006 & 0.005 & 1.05 \\
Tg & \emph{T. gondii} & GCF\_000006565.2 & 0.067 & 0.065 & 0.062 & 1.06 \\
Mb & \emph{M. balbisiana} & GCA\_004837865.1 & 0.50 & 0.49 & 0.37 & 1.33 \\
Hs38 & Human GRCh38 & GCF\_000001405.40 & 3.34 & 3.14 & 2.51 & 1.25 \\
HsT2T & Human CHM13 & GCF\_009914755.1 & 3.16 & 3.12 & 2.51 & 1.24 \\
Pt & \emph{Pinus taeda} & GCA\_000404065.3 & 22.5 & 20.4 & 10.5 & 1.95 \\
CoV & SARS-CoV-2 (620k) & AGC~\cite{agc} & 18.8 & 18.2 & 0.002 & 9\,393 \\
Ec-r & \emph{E. coli} reads & SRR2584863 & 1.14 & 0.37 & 0.045 & 8.32 \\
Ce-r & \emph{C. elegans} reads & SRR065390 & 17.6 & 4.72 & 0.49 & 9.68 \\
Hs-r & Human reads & ERR3239334 & 278 & 91.0 & 6.46 & 14.1 \\
\bottomrule
\end{tabular}
\end{table}

To verify the results, we compare every \mkc{} index exactly with the \kmc{} database of the same input. We look up each \kmer{} listed by \kmc{} through the public interface of \mkc, mark the record that holds it in a bitmap, which exposes duplicates, and check that the number of marked records equals the number of occupied records. All \VerifiedTotal{} comparisons, covering every data set at $k=31$ and the sweeps of Section~\ref{sec:ksweep} (\ExactKmers{} billion \kmers), succeeded. In addition, the query benchmarks check every returned count.

\subsection{Index Construction}\label{sec:construction}
\begin{table*}[t]
\caption{Building a queryable index of all canonical 31-mers with 16 threads: time until the index is durable (median of three runs $\pm$ half their range), peak resident memory and size of the index on disk. Best value per row in bold. J: \jf, H: \hg. $\dagger$: \kmc{} was faster with \texttt{-m4} than with \texttt{-m16}; time and memory are those of the faster setting.}
\label{tab:build}
\centering\footnotesize\setlength{\tabcolsep}{4pt}
\begin{tabular}{@{}l llll c rrrr c rrrr@{}}
\toprule
 & \multicolumn{4}{c}{Time (s)} && \multicolumn{4}{c}{Peak memory (GiB)} && \multicolumn{4}{c}{Index size (GB)}\\
\cmidrule{2-5}\cmidrule{7-10}\cmidrule{12-15}
 & \mkc & \kmc & J & H && \mkc & \kmc & J & H && \mkc & \kmc & J & H\\
\midrule
Ec & 0.18{\tiny$\,\pm$.00} & 1.12$^{\dagger}${\tiny$\,\pm$.02} & \textbf{0.16}{\tiny$\,\pm$.00} & 7.83{\tiny$\,\pm$.03} &  & \textbf{0.076} & 0.12 & 0.077 & 0.36 &  & \textbf{0.033} & 0.044 & 0.064 & 0.040 \\
Tg & \textbf{0.46}{\tiny$\,\pm$.01} & 1.95$^{\dagger}${\tiny$\,\pm$.02} & 1.67{\tiny$\,\pm$.01} & 9.20{\tiny$\,\pm$.09} &  & \textbf{0.16} & 0.93 & 0.82 & 0.77 &  & \textbf{0.33} & 0.56 & 0.74 & 0.49 \\
Mb & \textbf{1.81}{\tiny$\,\pm$.01} & 6.13$^{\dagger}${\tiny$\,\pm$.03} & 10.5{\tiny$\,\pm$0.12} & 17.5{\tiny$\,\pm$0.67} &  & \textbf{0.30} & 3.60 & 3.10 & 3.24 &  & \textbf{1.95} & 3.70 & 4.44 & 3.13 \\
Hs38 & \textbf{9.05}{\tiny$\,\pm$.01} & 25.1$^{\dagger}${\tiny$\,\pm$0.46} & 74.1{\tiny$\,\pm$0.21} & 68.2{\tiny$\,\pm$1.45} &  & \textbf{0.94} & 3.76 & 23.1 & 18.1 &  & \textbf{13.3} & 25.1 & 30.1 & 19.1 \\
HsT2T & \textbf{8.86}{\tiny$\,\pm$.04} & 24.9$^{\dagger}${\tiny$\,\pm$0.69} & 73.6{\tiny$\,\pm$0.75} & 70.5{\tiny$\,\pm$3.16} &  & \textbf{1.01} & 3.76 & 23.1 & 18.5 &  & \textbf{11.7} & 25.1 & 30.1 & 19.5 \\
Pt & \textbf{73.9}{\tiny$\,\pm$5.15} & 122{\tiny$\,\pm$1.56} & 458{\tiny$\,\pm$2.85} & 412{\tiny$\,\pm$12.3} &  & \textbf{4.13} & 14.9 & 88.0 & 76.4 &  & \textbf{55.8} & 105 & 126 & 81.7 \\
CoV & \textbf{14.8}{\tiny$\,\pm$.02} & 118{\tiny$\,\pm$0.34} & 109{\tiny$\,\pm$0.33} & 287{\tiny$\,\pm$2.11} &  & 0.28 & 14.8 & \textbf{0.047} & 0.34 &  & \textbf{0.015} & 0.021 & 0.023 & 0.018 \\
Ec-r & \textbf{1.23}{\tiny$\,\pm$.01} & 3.26$^{\dagger}${\tiny$\,\pm$.02} & 4.36{\tiny$\,\pm$0.20} & 13.0{\tiny$\,\pm$.04} &  & \textbf{0.28} & 3.38 & 0.43 & 0.66 &  & \textbf{0.30} & 0.45 & 0.54 & 0.37 \\
Ce-r & \textbf{9.47}{\tiny$\,\pm$.01} & 21.9$^{\dagger}${\tiny$\,\pm$0.12} & 60.4{\tiny$\,\pm$3.46} & 80.2{\tiny$\,\pm$2.84} &  & \textbf{1.41} & 3.75 & 6.02 & 4.03 &  & \textbf{3.02} & 4.46 & 5.85 & 4.00 \\
Hs-r & \textbf{209}{\tiny$\,\pm$2.96} & 492{\tiny$\,\pm$7.86} & 786{\tiny$\,\pm$43.4} & 1\,450{\tiny$\,\pm$14.5} &  & \textbf{5.14} & 15.0 & 45.2 & 53.0 &  & \textbf{36.7} & 64.7 & 77.5 & 56.5 \\
\bottomrule
\end{tabular}
\end{table*}

As Table~\ref{tab:build} shows, \mkc{} is the fastest tool on every input except the 5\,Mbp \emph{E.~coli} genome, on which Jellyfish, which writes no temporary files, finishes first. \mkc{} is \BuildSpeedKmcMin--\BuildSpeedKmcMax\x{} faster than the better \kmc{} setting. For example, for the human genome (HsT2T) \mkc{} takes 8.9\,s, \kmc{} 24.9\,s, and the in-memory counters 70.5 and 73.6\,s. Half the range of the repetitions is at most 6\% of the median for every tool and data set except \mkc{} on Pt (7\%), where two of the runs read 13 and 19\,GB back from the SSD. On HsT2T, splitting takes one fifth of the time of \mkc, counting two fifths and assembly one third, and on Pt, assembly writes a 56\,GB index and takes almost half of the time, more than counting. The peak resident set size of \mkc{} is 1.0\,GiB for the human genome, about a quarter of the memory of \kmc{} with \texttt{-m4} and at most an eighteenth of that of Jellyfish and \hg. Its index is the smallest of the four, i.e., \SizeRatioKmcMin--\SizeRatioKmcMax\x{} smaller than the database of \kmc, \SizeRatioJfMin--\SizeRatioJfMax\x{} smaller than that of Jellyfish and \SizeRatioHgMin--\SizeRatioHgMax\x{} smaller than that of \hg.

All input pages were resident at the start of every build except on Hs-r, whose 278\,GB cannot be cached (41--45\% were resident) and were read from the SSD by every tool. When we evict the input before each build, \mkc{} still takes 8.9\,s on HsT2T and 12.8 instead of 9.5\,s on Ce-r, and \kmc{} with \texttt{-m16} 28.1 instead of 28.4\,s and 29.0 instead of 25.6\,s, so the ranking does not change. On a slower device, however, reading the input would dominate the time of both tools.

\subsection{Index Anatomy}\label{sec:anatomy}
\begin{table}[t]
\caption{Structure of the \mkc{} hash index ($k=31$). Shape: slots\x{}record bits, or $S$\,EF for packed buckets of up to $S$ keys (Section~\ref{sec:packed}), for which $t$ is the tag width of the count list; $q$, $t$: quotient and tag bits; b/k: bits per distinct \kmer{} for the whole file; 1st: share of \kmers{} and of occurrences stored in their first-choice bucket; filt.: share of buckets with an overflow-filter bit set; large: share of \kmers{} whose count is not stored in the record; lines: mean number of cache lines loaded per lookup, measured by an instrumented reader, for stored (pr.), input-order (nat.) and absent (abs.) \kmers.}
\label{tab:anatomy}
\centering\footnotesize\setlength{\tabcolsep}{2.4pt}
\begin{tabular}{@{}lcrrrrrrrrrrr@{}}
\toprule
 & Shape & $q$ & $t$ & b/k & \multicolumn{2}{c}{1st (\%)} & filt. & large & stash & \multicolumn{3}{c}{lines}\\
 & & & & & keys & occ. & (\%) & (\%) & & pr. & nat. & abs.\\
\midrule
Ec & 11\,EF & 43 & 7 & 49.5 & 90.1 & 90.5 & 36.2 & 0.000 & 0 & 1.10 & 1.10 & 1.03 \\
Tg & 13\,EF & 39 & 12 & 42.7 & 90.9 & 91.3 & 35.6 & 0.000 & 10 & 1.09 & 1.09 & 1.17 \\
Mb & 13\,EF & 37 & 10 & 42.1 & 91.2 & 92.9 & 35.2 & 0.004 & 0 & 1.09 & 1.11 & 1.05 \\
Hs38 & 13\x39 & 34 & 4 & 42.5 & 91.3 & 92.8 & 35.2 & 0.159 & 0 & 1.09 & 1.30 & 1.16 \\
HsT2T & 15\,EF & 34 & 7 & 37.4 & 91.8 & 93.2 & 34.8 & 0.033 & 813 & 1.08 & 1.17 & 1.15 \\
Pt & 13\x39 & 32 & 6 & 42.5 & 91.3 & 94.9 & 35.1 & 0.156 & 0 & 1.09 & 1.26 & 1.16 \\
CoV & 9\x56 & 44 & 11 & 62.5 & 88.3 & 96.3 & 37.9 & 3.126 & 0 & 1.18 & 3.03 & 1.11 \\
Ec-r & 10\x51 & 39 & 11 & 54.5 & 89.3 & 96.1 & 37.0 & 0.006 & 0 & 1.11 & 1.11 & 1.28 \\
Ce-r & 11\x46 & 36 & 9 & 49.6 & 90.1 & 96.7 & 36.2 & 0.052 & 0 & 1.10 & 1.20 & 1.13 \\
Hs-r & 12\x42 & 32 & 9 & 45.5 & 90.7 & 96.7 & 35.6 & 0.105 & 0 & 1.09 & 1.58 & 1.11 \\
\bottomrule
\end{tabular}
\end{table}

Table~\ref{tab:anatomy} lists the structure that \mkc{} chooses for each index. Genomes have few large counts and get short tags and more slots, e.g., thirteen 39-bit slots with a four-bit tag for Hs38 (\BitsHuman{} bits per \kmer), and HsT2T and the three smaller genomes get packed buckets (\BitsHumanPacked{} bits per \kmer{} for HsT2T, 13.7 keys per bucket on average). The read sets and the SARS-CoV-2 collection have many large counts and get tags of nine to eleven bits and nine to twelve slots. About nine keys in ten are stored in their first-choice bucket.

An instrumented reader counts every access to a bucket, a group offset, the count array and the dictionary. A stored \kmer{} costs 1.08--1.11 cache lines (1.18 on CoV). \kmers{} in input order cost 1.17--1.30 lines on the large genomes, because repeated \kmers{} are over-represented among them and 19--24\% of these lookups on Hs38 and Pt read a count from the dictionary, and they cost 3.03 lines on CoV, where nearly every \kmer{} occurs in thousands of genomes. An absent \kmer{} costs 1.03--1.17 lines on genomes and 1.11--1.28 lines on read sets. Although a third of the buckets have displaced a key, the filter stops most misses, so that only 16\% of the absent lookups on Hs38 and 7\% on HsT2T reach a second bucket.

These indexes also test the assumption of Section~\ref{sec:index}. For independent uniform hashes, the group sizes would follow a binomial distribution. On HsT2T, the standard deviation of the group sizes is 138.6 against 138.4 in the binomial model, the fullest group is at 94.6\% of its capacity, and on the other data sets the two values agree within 2\%.

\subsection{Point Queries}\label{sec:query}
\begin{table*}[t]
\caption{Single-thread point-query throughput in million lookups per second ($k=31$). All tools are compared on 4\,KiB pages (file mapping for \mkc, plain heap for the others) and on 2\,MiB pages. \emph{present}: uniform over stored \kmers; \emph{absent}: uniformly random \kmers{} not in the index; \emph{seq.}: consecutive \kmers{} of the input, taken from 64 places spread evenly over it. Medians of six samples; best value in bold. H: \hg, J: \jf, K: \kmc. Last columns: the batch interface of \mkc{} on 2\,MiB pages. Absent \kmers{} on 2\,MiB pages are omitted for space; the text gives their ratios. Patterns of 16 million \kmers{} except on Ec and CoV (one million). --: Jellyfish on 2\,MiB pages ran out of memory on Pt.}
\label{tab:query}
\centering\footnotesize\setlength{\tabcolsep}{2.6pt}
\begin{tabular}{@{}l rrrr c rrrr c rrrr c rrrr c rrrr c rr@{}}
\toprule
 & \multicolumn{14}{c}{4\,KiB pages} && \multicolumn{9}{c}{2\,MiB pages} && \multicolumn{2}{c}{batch}\\
\cmidrule{2-15}\cmidrule{17-25}\cmidrule{27-28}
 & \multicolumn{4}{c}{present} && \multicolumn{4}{c}{absent} && \multicolumn{4}{c}{seq.} && \multicolumn{4}{c}{present} && \multicolumn{4}{c}{seq.} && pr. & seq.\\
\cmidrule{2-5}\cmidrule{7-10}\cmidrule{12-15}\cmidrule{17-20}\cmidrule{22-25}
 & \mkc & H & J & K && \mkc & H & J & K && \mkc & H & J & K && \mkc & H & J & K && \mkc & H & J & K && \multicolumn{2}{c}{\mkc}\\
\midrule
Ec & \textbf{32.5} & 15.8 & 3.20 & 4.23 &  & \textbf{34.3} & 10.5 & 2.93 & 4.56 &  & \textbf{32.3} & 16.8 & 3.17 & 5.09 &  & \textbf{32.8} & 18.2 & 3.37 & 4.52 &  & \textbf{32.4} & 18.8 & 3.39 & 5.47 &  & 83.9 & 85.9 \\
Tg & \textbf{14.6} & 7.95 & 2.16 & 2.39 &  & \textbf{14.9} & 5.68 & 2.04 & 2.43 &  & \textbf{14.9} & 9.07 & 2.20 & 2.70 &  & \textbf{16.1} & 8.85 & 2.34 & 2.57 &  & \textbf{16.6} & 10.1 & 2.38 & 2.92 &  & 65.4 & 66.1 \\
Mb & \textbf{13.8} & 7.59 & 1.97 & 1.68 &  & \textbf{14.6} & 5.43 & 1.87 & 1.81 &  & \textbf{14.9} & 9.47 & 2.06 & 1.88 &  & \textbf{15.5} & 8.61 & 2.16 & 1.91 &  & \textbf{16.7} & 10.7 & 2.27 & 2.16 &  & 57.5 & 61.3 \\
Hs38 & \textbf{12.8} & 6.06 & 1.66 & 1.12 &  & \textbf{12.2} & 4.29 & 1.58 & 1.26 &  & \textbf{12.6} & 6.90 & 1.67 & 1.25 &  & \textbf{15.5} & 8.11 & 2.11 & 1.48 &  & \textbf{15.7} & 9.46 & 2.15 & 1.63 &  & 76.6 & 71.9 \\
HsT2T & \textbf{13.0} & 6.10 & 1.63 & 1.12 &  & \textbf{13.3} & 4.32 & 1.57 & 1.26 &  & \textbf{12.7} & 7.27 & 1.70 & 1.29 &  & \textbf{15.0} & 8.19 & 2.06 & 1.50 &  & \textbf{15.8} & 10.2 & 2.16 & 1.70 &  & 73.4 & 73.2 \\
Pt & \textbf{8.98} & 4.73 & 1.48 & 0.89 &  & \textbf{8.67} & 3.34 & 1.42 & 0.97 &  & \textbf{9.16} & 6.22 & 1.49 & 0.94 &  & \textbf{14.9} & 6.16 & -- & 1.06 &  & \textbf{14.9} & 7.90 & -- & 1.12 &  & 75.9 & 66.9 \\
CoV & \textbf{50.7} & 23.8 & 5.34 & 5.91 &  & \textbf{57.8} & 17.4 & 4.79 & 6.40 &  & 46.3 & \textbf{46.5} & 5.52 & 8.12 &  & \textbf{50.8} & 26.9 & 5.78 & 6.24 &  & 46.3 & \textbf{51.3} & 5.80 & 8.31 &  & 90.5 & 84.1 \\
Ec-r & \textbf{15.1} & 8.08 & 2.21 & 2.43 &  & \textbf{12.8} & 5.76 & 2.08 & 2.57 &  & \textbf{17.7} & 14.9 & 2.28 & 2.81 &  & \textbf{16.9} & 8.94 & 2.38 & 2.58 &  & \textbf{20.0} & 16.5 & 2.45 & 2.99 &  & 69.3 & 90.9 \\
Ce-r & \textbf{14.2} & 7.50 & 1.98 & 2.17 &  & \textbf{14.0} & 5.36 & 1.89 & 2.50 &  & \textbf{15.7} & 12.6 & 2.01 & 2.39 &  & \textbf{16.2} & 8.49 & 2.15 & 2.42 &  & \textbf{18.2} & 14.3 & 2.19 & 2.63 &  & 64.7 & 72.7 \\
Hs-r & \textbf{9.63} & 4.83 & 1.52 & 1.41 &  & \textbf{9.69} & 3.43 & 1.45 & 1.59 &  & \textbf{41.4} & 29.8 & 4.28 & 5.59 &  & \textbf{15.3} & 7.09 & 1.79 & 1.56 &  & \textbf{53.5} & 35.7 & 4.60 & 6.20 &  & 77.3 & 106 \\
\bottomrule
\end{tabular}
\end{table*}

Since the page size alone changes single-thread throughput by up to 65\%, we compare the tools at equal page size. On 4\,KiB pages, \mkc{} uses a file mapping and the other tools run unchanged. On 2\,MiB pages, \mkc{} uses its in-memory copy, and the other tools use a heap backed by transparent huge pages through the \texttt{hugetlb} tunable of the allocator or, for \hg, explicit advice. Each query pattern holds 16 million \kmers{} and is queried in one pass, except on Ec and CoV, whose indexes fit in the cache, where a pattern holds one million \kmers{} and is queried in eight passes by \mkc{} and two by the other tools. The kernel grants 2\,MiB pages only as far as it finds contiguous free memory. On Mb, Hs38, HsT2T and Ce-r, at least 95\% of the resident memory of \mkc, \kmc{} and \hg{} was on 2\,MiB pages. On Pt and Hs-r, it was all of the index of \mkc, but 48--58\% for \kmc{} and 4--84\% for \hg, and 19--35\% for Jellyfish on Hs-r against 41--49\% on the human genomes, so the 2\,MiB columns of these two rows favour \mkc; the 4\,KiB columns are not affected.

\textbf{Independent lookups.} For random stored and absent \kmers, \mkc{} is \LikeSmallHgRandomMin--\LikeSmallHgRandomMax\x{} faster than \hg{} on 4\,KiB pages and \LikeHugeHgRandomMin--\LikeHugeHgRandomMax\x{} faster on 2\,MiB pages, \LikeSmallJfRandomMin--\LikeSmallJfRandomMax\x{} and \LikeHugeJfRandomMin--\LikeHugeJfRandomMax\x{} faster than Jellyfish, and \LikeSmallKmcRandomMin--\LikeSmallKmcRandomMax\x{} and \LikeHugeKmcRandomMin--\LikeHugeKmcRandomMax\x{} faster than \kmc. Because of the overflow filter, an absent \kmer{} costs \mkc{} about as much time as a stored one (between 12\% less and 18\% more), whereas \hg{} takes at least 37\% more time for absent \kmers. \kmers{} that differ from a stored \kmer{} by one substitution are answered at the speed of random absent \kmers.

\textbf{\kmers{} in sequence order.} For \kmers{} in sequence order, the advantage of \mkc{} is smaller, because \hg{} is 1.0--1.3\x{} faster on them than on random stored \kmers{} on genomes and 1.7--2.0\x{} on Ec-r, Ce-r and CoV, whereas \mkc{} gains at most 18\%; on Hs-r, both tools gain severalfold (\hg{} 5.0--6.2\x, \mkc{} 3.5--4.3\x). As a result, \mkc{} is \LikeSmallHgSpreadMin--\LikeSmallHgSpreadMax\x{} as fast as \hg{} on 4\,KiB pages and \LikeHugeHgSpreadMin--\LikeHugeHgSpreadMax\x{} as fast on 2\,MiB pages, with the smallest factors on Ec-r and Ce-r (1.2--1.3) and on CoV, whose index fits in the cache and on which \hg{} is 10\% faster on 2\,MiB pages. For the first \kmers{} of the input instead of \kmers{} from 64 places spread over it, the factor on 2\,MiB pages is only \LikeHugeHgNaturalMin--\LikeHugeHgNaturalMax, and we have not established the reason. A batch interface can exploit that neighbouring \kmers{} arrive together. The batch interface of \mkc{} resolves 64 \kmers{} at a time with prefetching and answers 61--106 million \kmers{} in sequence order per second on one thread, \BatchHgSpreadMin--\BatchHgSpreadMax\x{} more than \hg.

\textbf{Threads, memory and latency.} With 16 threads, \mkc{} is \LikeHugeHgRandomMTMin--\LikeHugeHgRandomMTMax\x{} as fast as \hg{} on 2\,MiB pages for random \kmers{} and \LikeHugeHgSpreadMTMin--\LikeHugeHgSpreadMTMax\x{} as fast in sequence order, e.g., 189 against 104 million random lookups per second on HsT2T, and its batch interface answers 341 million. While the HsT2T index is queried, the peak resident set size of the process is 11.0\,GiB for \mkc, 18.4\,GiB for \hg, 23.5\,GiB for \kmc{} and 28.2\,GiB for Jellyfish. Loading the copy on 2\,MiB pages takes 2.7--6.0\,s and pays off after 260--590 million single-thread lookups, but the copy is private to each process: four processes hold 43.8\,GiB in the default mode and share 10.5\,GiB when they map the file. Timed individually over one million uncached \kmers{} of HsT2T, the median lookup takes 110\,ns for \mkc{} on 2\,MiB pages, 190\,ns on 4\,KiB pages, 630\,ns for Jellyfish and 925\,ns for \kmc, with 99.9th percentiles of 525, 640, 1\,210 and 2\,160\,ns.

\subsection{Comparison with a Compressed Dictionary}\label{sec:sshash}
\begin{table}[t]
\caption{\mkc{} and weighted SSHash (S) on four genomes, $k=31$. Bits per distinct \kmer{} including counts; time to build from sequence with 16 threads (S: unitig construction and reordering, then index construction); single-thread lookups in millions per second on 4\,KiB pages; seq.: the first million consecutive \kmers{} of the input, for which SSHash also offers streaming queries (str.).}
\label{tab:sshash}
\centering\footnotesize\setlength{\tabcolsep}{2.0pt}
\begin{tabular}{@{}l rr rrr rr rr rrr@{}}
\toprule
 & \multicolumn{2}{c}{bits/\kmer} & \multicolumn{3}{c}{build (s)} & \multicolumn{2}{c}{present} & \multicolumn{2}{c}{absent} & \multicolumn{3}{c}{seq.}\\
\cmidrule(lr){2-3}\cmidrule(lr){4-6}\cmidrule(lr){7-8}\cmidrule(lr){9-10}\cmidrule(l){11-13}
 & \mkc & S & \mkc & \multicolumn{2}{c}{S} & \mkc & S & \mkc & S & \mkc & S & str.\\
\midrule
Ec & 49.5 & 5.3 & 0.18 & 4.79 & 0.29 & 32.6 & 6.97 & 35.0 & 11.9 & 32.2 & 9.57 & 47.0 \\
Tg & 42.7 & 6.2 & 0.46 & 32.8 & 2.69 & 16.7 & 4.56 & 17.1 & 6.83 & 16.9 & 9.32 & 46.0 \\
Mb & 42.1 & 8.2 & 1.81 & 289 & 17.6 & 14.3 & 2.11 & 16.0 & 3.67 & 15.8 & 7.88 & 32.0 \\
HsT2T & 37.4 & 9.1 & 8.86 & 1\,791 & 132 & 13.2 & 1.50 & 13.5 & 2.96 & 13.7 & 5.51 & 14.4 \\
\bottomrule
\end{tabular}
\end{table}

Weighted SSHash~\cite{sshashw} stores the same information as the index of \mkc, i.e., the exact count of every \kmer{} of a fixed set, but represents the \kmers{} as overlapping strings. We build it for four genomes with its documented pipeline and 16 threads: BCALM\,2~\cite{bcalm2} computes unitigs with abundances, UST~\cite{ust} joins them, and the current SSHash~\cite{sshashopt} reorders and indexes them with the advised minimizer lengths of 16 to 21.

As Table~\ref{tab:sshash} shows, SSHash needs 5.3--9.1 bits per \kmer{} including its count and \mkc{} 37.4--49.5 bits, e.g., 2.9 against 11.7\,GB for the human genome, so for applications in which the size of the index is the main constraint, a compressed dictionary is the better choice. In contrast, the SSHash pipeline needs 32\,min, 16\,GiB of memory and 89\,GB of scratch space for the human genome, against 8.9\,s and 1.0\,GiB for \mkc, and the ratio of the build times exceeds 100 on all but the two smallest genomes. \mkc{} also answers lookups of stored \kmers{} 3.7--8.8\x{} and of absent \kmers{} 2.5--4.6\x{} faster, because an SSHash lookup evaluates a minimal perfect hash and then reads and compares string data. For \kmers{} in sequence order, the streaming mode of SSHash, which extends the previous match along the stored string, processes 14.4 million \kmers{} per second on HsT2T, slightly more than single lookups of \mkc{} (13.7 million) in a fourth of the memory, while the batch interface of \mkc{} processes 76 million.

\subsection{Ablation Study}\label{sec:ablation}
\begin{table}[t]
\caption{Variants of the HsT2T index built by the same builder: nominal load (default 94\%) and placement in arrival order instead of by count, with packed buckets; then the record layout of the same keys (default shape: thirteen 39-bit records with four-bit tags), its shapes with tags of at least seven bits only, a single overflow flag instead of the filter, and a reader that ignores the filter. 1st: share of keys / occurrences in the first-choice bucket; filt.: share of buckets with a filter bit set; lines: cache lines per lookup; M/s: single-thread throughput on 2\,MiB pages for stored, absent and the first million consecutive \kmers.}
\label{tab:ablate}
\centering\footnotesize\setlength{\tabcolsep}{1.3pt}
\begin{tabular}{@{}lrrrrrrrrrrr@{}}
\toprule
 & b/k & build & \multicolumn{2}{c}{1st (\%)} & filt. & stash & \multicolumn{2}{c}{lines} & \multicolumn{3}{c}{M/s}\\
 & & (s) & keys & occ. & (\%) & & pr. & abs. & pr. & abs. & seq.\\
\midrule
80\% & 43.9 & 8.41 & 96.4 & 97.1 & 16.0 & 29 & 1.04 & 1.05 & 17.0 & 18.2 & 18.3 \\
85\% & 41.3 & 8.49 & 95.0 & 96.0 & 21.8 & 139 & 1.05 & 1.08 & 17.3 & 17.9 & 18.1 \\
90\% & 39.0 & 8.61 & 93.4 & 94.6 & 28.5 & 342 & 1.07 & 1.11 & 17.0 & 17.3 & 17.9 \\
94\% (default) & 37.4 & 8.78 & 91.8 & 93.2 & 34.8 & 813 & 1.08 & 1.15 & 16.8 & 17.2 & 17.8 \\
96\% & 36.6 & 8.86 & 90.7 & 92.2 & 39.4 & 1\,099 & 1.09 & 1.16 & 16.6 & 17.1 & 16.7 \\
97\% & 36.2 & 8.95 & 89.8 & 91.3 & 43.4 & 1\,449 & 1.10 & 1.20 & 16.3 & 16.6 & 17.7 \\
98\% & 35.8 & 9.13 & 88.6 & 90.1 & 49.6 & 7\,308 & 1.12 & 1.22 & 15.9 & 16.5 & 17.1 \\
\midrule
arrival order & 37.7 & 9.61 & 91.9 & 91.2 & 35.1 & 1.57\,M & 1.08 & 1.15 & 16.7 & 16.6 & 16.4 \\
\midrule
\multicolumn{12}{@{}l}{\emph{record buckets, same keys:}}\\
default shape & 42.5 & 8.98 & 91.3 & 92.8 & 35.1 & 0 & 1.09 & 1.16 & 17.1 & 16.4 & 18.1 \\
tags $\ge 7$ bits & 45.4 & 8.53 & 90.7 & 92.3 & 35.6 & 0 & 1.09 & 1.11 & 17.3 & 17.8 & 18.0 \\
one overflow flag & 42.5 & 8.96 & 91.3 & 92.8 & 35.1 & 0 & 1.09 & 1.35 & 16.9 & 13.0 & 18.1 \\
flags ignored & 42.5 & 8.98 & 91.3 & 92.8 & 35.1 & 0 & 1.09 & 1.71 & 17.3 & 8.84 & 18.0 \\
\bottomrule
\end{tabular}
\end{table}

To analyze the contribution of each design decision, we build variants of the HsT2T index from the same commit with one decision changed through a build-time switch (Table~\ref{tab:ablate}), and we quote Ce-r where it differs. All rows use the same fixtures of one million keys. These variants cover the decisions of the index and its reader; those of construction are measured where they are described, i.e., the minimizer order against a hash order (Section~\ref{sec:split}), the hash seed (Section~\ref{sec:robust}) and the workspace budget (Section~\ref{sec:memory}). The offline assembly itself has no variant with one switch changed: placement by count, the fixed load and exact overflow filters all require the complete key set of a group before its first key is placed, so a builder that inserts keys while counting them cannot produce this index, and the comparison with the counters of Section~\ref{sec:construction}, which insert while counting or write sorted lists, takes the place of such a variant.

\textbf{Load.} From 80\% to 96\% load, the packed HsT2T index shrinks from 43.9 to 36.6 bits per \kmer, the share of buckets with a filter bit set rises from 16\% to 39\%, and throughput drops by 2\% for stored and 6\% for absent keys, while the build time increases by 9\% up to 98\%. A few keys fit into neither bucket at every load (813 at the default), and with record buckets, placement first fails at 97\% load on Ce-r. The default of 94\% thus leaves a margin of three percentage points.

\textbf{Placement by count.} Placing keys by decreasing count instead of in arrival order raises the share of occurrences in the first-choice bucket from 91.2\% to 93.2\% on HsT2T and from 90.1\% to 96.7\% on Ce-r, and the throughput in sequence order by 9\% and 10\%, at no cost in build time. For packed buckets, it also bounds the stash, which in arrival order holds 1.57 million HsT2T keys whose count lists are full.

\textbf{Packed buckets and short tags.} On the same keys, the record layout takes \BitsHumanRecords{} instead of \BitsHumanPacked{} bits per \kmer, builds 2\% slower and answers stored \kmers{} 2\% faster but absent \kmers{} 5\% slower. Restricting the record layout to tags of at least seven bits costs another 7\% of space, but makes absent lookups 9\% faster, because twelve 42-bit slots leave eight filter bits instead of five.

\textbf{Overflow filter.} With one overflow flag per bucket instead of the filter, a lookup of an absent \kmer{} loads 1.35 instead of 1.16 cache lines and throughput drops by 21\% (23\% on Ce-r). With a reader that ignores the filter, it loads 1.71 lines and throughput drops by 46\%, while lookups of stored \kmers{} do not change in either case. Of all design decisions, the filter thus has the largest effect, and because the table is static, its bits are never stale.

\textbf{Kernels and batching.} The specialized AVX2 kernels are 1.4--1.8\x{} faster than the portable scalar path, and the batch interface is about five times faster than single lookups (73 against 15.0 million stored keys per second on HsT2T with 16 million keys).

\subsection{Robustness to the Key Set}\label{sec:robust}
\begin{table}[t]
\caption{Key sets of 20 million 31-mers built with 16 threads, without and with reseeding (the seed finally used in parentheses). Build: total time, including any abandoned attempt; b/k: bits per \kmer; lines: per lookup of a stored \kmer; M/s: single-thread lookups of stored \kmers{}; exact: listing equal to \kmc's.}
\label{tab:adversarial}
\centering\footnotesize\setlength{\tabcolsep}{1.1pt}
\begin{tabular}{@{}llrrrrrc@{}}
\toprule
Key set & reseed & build (s) & stash & b/k & lines & M/s & exact\\
\midrule
random hashes & off & 0.83 & 0 & 45.4 & 1.09 & 21.2 & yes \\
random hashes & on (0) & 0.83 & 0 & 45.4 & 1.09 & 21.4 & yes \\
1\,M keys in one group & off & 5.72 & 998\,k & 65.2 & 3.08 & 29.4 & yes \\
1\,M keys in one group & on (1) & 3.27 & 0 & 45.4 & 1.09 & 21.3 & yes \\
all in 1/1024 of range & off & 99.8 & 19.98\,M & 602.8 & 3.05 & 12.5 & yes \\
all in 1/1024 of range & on (1) & 6.87 & 0 & 45.4 & 1.09 & 21.2 & yes \\
\bottomrule
\end{tabular}
\end{table}

Since $\pi_b$ is a fixed bijection, adverse key sets can be constructed by inverting it. We generate two sets of 20 million distinct 31-mers: in ``1\,M keys in one group'', one million keys share their top 30 hash bits and thus fall into one group for every shape, and in ``all in 1/1024 of range'', all keys fall into one spool partition. Every 997th key occurs three times and every 9\,973rd key 70\,000 times. With reseeding disabled, \mkc{} remains exact and stores the overflow in the stash, i.e., 998\,000 keys for the first set and nearly all keys for the second, whose index takes 603 bits per \kmer{} and 100\,s to build (Table~\ref{tab:adversarial}). By default, \mkc{} detects the overflow and builds again with the next seed, and both sets give an ordinary index (empty stash, 45.4 bits per \kmer) in 3.3 and 6.9\,s including the abandoned attempt. Reseeding is not a guarantee, however, since key sets can be chosen against all four functions, and it requires inputs that can be read again. On real data, the seed has no measurable effect: with each of the four seeds, the HsT2T and Ce-r indexes have the same size, first-bucket share and filter occupancy, and build and lookup times differ by at most 1\% and 6\%.

\subsection{Application to Assembly Evaluation}\label{sec:app}
To evaluate \mkc{} in an application, we compute the consensus quality value of Merqury~\cite{merqury}, which counts the \kmers{} of an assembly that are absent from the reads, on top of each index. With \mkc, a program of about 80 lines looks up the \kmers{} of the assembly through the batch interface with 16 threads. With \kmc, we count the assembly and subtract the read database with \texttt{kmc\_tools}, the intended way to combine sorted databases. For Ce-r against the \emph{C.~elegans} reference, both find the same 92\,932 missing \kmers{} (quality value 45.2), and given the read index the evaluation takes 1.8\,s with \mkc{} and 9.8\,s with \kmc, or 11.2 against 31.6\,s from reads to result. For Hs-r against HsT2T (78.5 million missing \kmers, quality value 30.8), it takes 24 against 77\,s (medians of three runs, which took 14.5, 24.3 and 24.3\,s with \mkc), of which \mkc{} spends 18\,s loading its 37\,GB index, and 233 against 569\,s from reads to result. This experiment measures only the step for which a lookup index is suited, and the other outputs of Merqury require sorted iteration and set operations, which \mkc{} does not provide.

\subsection{Effect of $k$}\label{sec:ksweep}
\begin{figure*}[t]
\centering
\begin{tikzpicture}
\begin{groupplot}[group style={group size=4 by 1, horizontal sep=1.0cm}, paper, width=0.27\textwidth, height=4.0cm, xlabel={$k$}, xtick={5,9,13,17,21,25,32}, xticklabels={5,9,13,17,21,25,32}, xmin=4, xmax=33]
\nextgroupplot[title={(a) build time (s)}, ymode=log, log ticks with fixed point, legend to name=kleg]
\addplot[smkc] coordinates {
  (5,0.8022049684077501)
  (7,0.7735515050590038)
  (9,1.0030052438378334)
  (11,1.741401532664895)
  (13,3.784432366490364)
  (14,6.123072100803256)
  (15,6.827821725979447)
  (16,7.3828288074582815)
  (17,7.821306215599179)
  (19,7.789216473698616)
  (21,8.293783102184534)
  (25,8.437909875065088)
  (29,8.583953412249684)
  (31,8.85589262470603)
  (32,8.861943555995822)
}; \addlegendentry{\mkc}
\addplot[skmc] coordinates {
  (5,3.9087111968547106)
  (7,3.8722831066697836)
  (9,3.916810940951109)
  (11,7.146018145605922)
  (13,12.594058685004711)
  (14,16.287407472729683)
  (15,16.963398404419422)
  (16,17.901972515508533)
  (17,20.057595623657107)
  (19,22.29814720712602)
  (21,22.798372205346823)
  (25,23.25764887779951)
  (29,25.4559117089957)
  (31,24.87879029661417)
  (32,27.692164923995733)
}; \addlegendentry{\kmc}
\addplot[sjf] coordinates {
  (5,26.07293713837862)
  (7,36.128803215920925)
  (9,44.59194799885154)
  (11,48.262097142636776)
  (13,118.93110871315002)
  (14,280.91481521725655)
  (15,29.841173000633717)
  (16,37.59568855538964)
  (17,44.1411757171154)
  (19,52.90472849458456)
  (21,58.06950192525983)
  (25,64.82795566692948)
  (29,70.7115893997252)
  (31,73.55886334180832)
  (32,71.20204824581742)
}; \addlegendentry{\jf}
\nextgroupplot[title={(b) index size (bits per distinct \kmer)}, ymin=0]
\addplot[smkc] coordinates {
  (5,35.0)
  (7,32.1875)
  (9,32.01171875)
  (11,16.251143126767026)
  (13,16.58470846942046)
  (14,21.215823038205833)
  (15,13.245441375170227)
  (16,38.911300961530515)
  (17,38.91483304599473)
  (19,38.91795573493646)
  (21,36.368260487560136)
  (25,36.344853117038)
  (29,36.337598141189375)
  (31,37.367257347645214)
  (32,41.93635438564346)
};
\addplot[skmc] coordinates {
  (5,41.8125)
  (7,32.58203125)
  (9,32.505126953125)
  (11,32.500333079172705)
  (13,32.51803846811748)
  (14,48.07827715605441)
  (15,48.03135996053126)
  (16,48.0209661396106)
  (17,48.03007385977844)
  (19,56.0050530905232)
  (21,56.018553105191394)
  (25,64.01753911740936)
  (29,72.0169342064621)
  (31,80.00417395844956)
  (32,80.00663485458738)
};
\addplot[sjf] coordinates {
  (5,56.75)
  (7,48.546875)
  (9,56.0341796875)
  (11,56.00213628445772)
  (13,64.0001383256544)
  (14,64.00003990391832)
  (15,64.00003445045327)
  (16,64.00001463554612)
  (17,72.00000862698029)
  (19,72.0000060136565)
  (21,80.00000569015963)
  (25,88.00000564679442)
  (29,96.00000573627034)
  (31,96.0000057825701)
  (32,96.00000582101362)
};
\nextgroupplot[title={(c) present queries, 1 thread (M/s)}, ymode=log, log ticks with fixed point]
\addplot[smkc] coordinates {
  (5,180.3867220170597)
  (7,181.3159795855538)
  (9,151.50251178476407)
  (11,158.50315012074543)
  (13,50.80447552232318)
  (14,42.65859310945276)
  (15,41.60345237874928)
  (16,18.139703085702934)
  (17,17.59609222868061)
  (19,17.056090065488462)
  (21,16.374829494846605)
  (25,16.40376220928404)
  (29,16.849568874884802)
  (31,15.033292460767477)
  (32,16.664776504751174)
};
\addplot[skmc] coordinates {
  (5,28.25826056416343)
  (7,27.660819528544174)
  (9,23.43522375097138)
  (11,17.70179185083284)
  (13,6.491212932498343)
  (14,2.4667489854466815)
  (15,2.0732735919018848)
  (16,1.7944340699916235)
  (17,1.6267610858456552)
  (19,1.3103166618213766)
  (21,1.4111169306229205)
  (25,1.407594380105289)
  (29,1.3077766884360509)
  (31,1.11592964753397)
  (32,1.1797632576308192)
};
\addplot[sjf] coordinates {
  (5,8.124178089729371)
  (7,4.950508726186124)
  (9,2.9247895045070527)
  (11,1.9487768409728474)
  (13,0.5012972947874745)
  (14,0.4196623219909978)
  (15,2.50957879214256)
  (16,2.420594754209958)
  (17,2.111232021771903)
  (19,1.8993441400590458)
  (21,1.8100711314245208)
  (25,1.6884063404726053)
  (29,1.6842351358465213)
  (31,1.625784649772089)
  (32,1.5963258685736326)
};
\nextgroupplot[title={(d) build time, $k=31$ (s)}, xmode=log, ymode=log, log ticks with fixed point, xtick={1,2,4,8,16,32}, xticklabels={1,2,4,8,16,32}, xmin=0.8, xmax=40, xlabel={threads}]
\addplot[smkc] coordinates {
  (1,107.90965336933732)
  (2,55.267697025090456)
  (4,28.556617360562086)
  (8,15.214474625885487)
  (16,8.811435893177986)
  (32,7.919081598520279)
};
\addplot[skmc] coordinates {
  (1,130.2021476365626)
  (2,78.67950062453747)
  (4,38.89854088053107)
  (8,27.40327099338174)
  (16,27.70027220621705)
  (32,28.768040388822556)
};
\addplot[sjf] coordinates {
  (1,nan)
  (2,nan)
  (4,237.1700922846794)
  (8,124.76522966846824)
  (16,72.60737227275968)
  (32,58.016029089689255)
};
\end{groupplot}
\node[anchor=south] at ($(group c2r1.north)!0.5!(group c3r1.north)+(0,0.45cm)$) {\pgfplotslegendfromname{kleg}};
\end{tikzpicture}
\caption{(a)--(c) Varying $k$ on HsT2T with 16 threads; \mkc{} writes the dense layout for $k\le\DenseMaxK$ and the hash layout above. (d) Varying the number of threads on HsT2T ($k=31$); the machine has 16 cores, and 32 threads use both hardware threads of every core.}
\label{fig:ksweep}
\end{figure*}
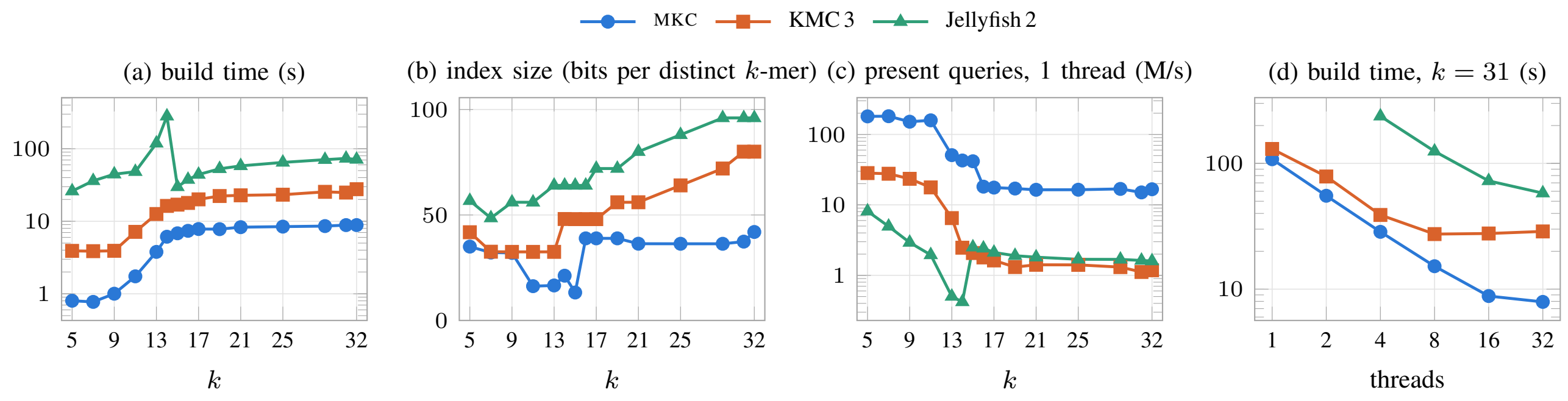

Fig.~\ref{fig:ksweep}(a)--(c) shows the effect of $k$ on HsT2T. \mkc{} is faster than \kmc{} for every $k$: 3.3--5.0\x{} for $k\le 13$, where it counts directly into an array that answers a lookup with one load, 2.5--2.7\x{} for $k=14$ and 15, where the sorting pipeline builds the dense layout, and 2.4--3.1\x{} from $k=16$ on (7.4--8.9\,s). The index takes 36.3--41.9 bits per \kmer{} for $16\le k\le 32$, whereas the sorted lists of \kmc{} and Jellyfish grow by two bits per base, to 80 and 96 bits at $k=31$, and the query throughput of the hash layout changes little with $k$ (15.0--18.1 million stored \kmers{} per second). For small $k$, the hash size passed to Jellyfish has no effect: for $k=11$, 13 and 14, it allocates $4^k$ table entries whether $n/0.8$ or a larger size is requested. At $k=14$, one of two such builds returned slightly too low counts for some highly repeated \kmers, and Fig.~\ref{fig:ksweep} shows the other, whose counts were correct for every queried \kmer. On Ce-r ($k=15$ to 25), \mkc{} is 2.0--2.3\x{} faster than \kmc.

\subsection{Scalability}\label{sec:scaling}
Fig.~\ref{fig:ksweep}(d) shows the effect of the number of threads on HsT2T. From one to 16 threads, the build time of \mkc{} drops by a factor of 12.2 (from 108 to 8.8\,s) and that of \kmc{} by a factor of 4.7 (from 130 to 27.7\,s), and \kmc{} does not become faster beyond eight threads. On one thread, \mkc{} takes 17\% less time than \kmc, although after sorting it also places every distinct \kmer{} into the cuckoo table and encodes its packed buckets, whereas \kmc{} only writes its sorted list. From eight threads on, \mkc{} is 1.8--3.6\x{} faster, the second hardware thread of each core makes it 11\% faster, and the index is byte-identical for all thread counts.

\subsection{Memory Budget and Disk Space}\label{sec:memory}
The workspace budget $M$ is the only resource parameter of \mkc. It limits the counting workspace and is not a bound on the resident set size. On HsT2T, lowering $M$ from 6400 to 400\,MiB makes \mkc{} use four times as many bins and the chunked path for 59 of them, and the build then takes 10.6 instead of 8.8\,s with a peak resident set size of 0.56 instead of 1.0\,GiB (Ce-r: 12.5 instead of 9.5\,s with 0.83 instead of 1.4\,GiB), which is more than this budget. Larger budgets change neither time nor resident set size. \kmc{} is fastest with \texttt{-m4} (24.2\,s on HsT2T within 3.8\,GiB), so \mkc{} with 400\,MiB is still faster than \kmc{} in every setting there. On Pt, a budget of 400\,MiB lengthens the build of \mkc{} from 69 to 124\,s in this sweep and lowers its peak from 4.1 to 1.5\,GiB, whereas \kmc{} takes 119--135\,s within 3.8--30\,GiB. Most of the extra time goes to assembly, because the 16 million exceptional counts no longer fit in the budget and their dictionary is built in the file, and since they are then left in the order in which the threads produced them, this is the only index of our experiments that differs between runs, in the order of these lists only.

The bins take about 8 bits per occurrence of a genome \kmer{} and the spool about as much as the index (13.3 against 11.7\,GB on HsT2T), so the temporary files of \mkc{} peak at 13.3\,GB on HsT2T and 54\,GB on Pt, against 3.0 and 20\,GB for \kmc, and at the same size as those of \kmc{} on Ce-r and Hs-r (5.1 and 95\,GB), where the bins dominate. Temporary files and index together take up to 25.0\,GB on HsT2T and 109\,GB on Pt, against 25.1 and 105\,GB for \kmc, whose database is about twice as large as the index of \mkc, whereas Jellyfish and \hg{} keep their intermediate data in memory. These sizes are sampled twice a second and can miss a shorter peak.

\subsection{Compressed Input}\label{sec:formats}
We build the Ce-r index from several encodings of the same 17.6\,GB of reads. Reading the gzip files as downloaded adds 3.3\,s to the 9.4\,s of \mkc{} and 6.0\,s to the 25.4\,s of \kmc, so \mkc{} counts the gzip files faster than \kmc{} counts the uncompressed files. Formats of independent blocks (BGZF, lz4) are decoded by all threads and add 1.0--1.7\,s, and zstd adds 5.1--5.3\,s, because its frames are decoded serially unless they declare their content size. All six indexes hold the same (\kmer, count) pairs as the \kmc{} database; for zstd and lz4 input, \mkc{} chooses 2048 instead of 1024 bins and writes a different file of the same size.

\section{Discussion}\label{sec:discussion}
\mkc{} should be read as a counter whose output is a static dictionary. Because the whole key set of a group is known before placement, \mkc{} can fill the buckets to 94\% load without long displacement chains, choose the width of the count field from the actual counts, place records by count, set exact overflow filters and lay out packed buckets. Of these decisions, the overflow filter has the largest effect on lookups, placement by count and short tags each change speed and space by a few percent, and packed buckets save an eighth of the space of a genome index. As a result, most lookups of stored and absent \kmers{} read one cache line, and construction makes random memory accesses only within the cache. On large inputs, it combines the small resident set of a disk-based counter with the shortest build time of the compared tools. For example, on 278\,GB of human reads, \mkc{} builds its index in 209\,s within 5.1\,GiB, \kmc{} in 492\,s within 15\,GiB, and the in-memory counters in 13 to 24 minutes within 45 to 53\,GiB.

\subsection{Limitations}
First, the index of \mkc{} is four to nine times larger than a compressed \kmer{} dictionary, which in turn takes one to two orders of magnitude longer to build and answers random lookups more slowly, so \mkc{} is not suitable when the size of the index is the main constraint. Second, its efficiency depends on the empirical observation that the hashes of real \kmers{} behave like random values. This observation held on every data set and seed that we tested, and reseeding handles key sets chosen against one hash function, but there is no worst-case guarantee. Third, in sequence order the advantage of single lookups over \hg{} is small (1.2--1.3\x{} on Ec-r and Ce-r, none on CoV), so applications that query whole sequences should use the batch interface. Fourth, the copy of the index on huge pages is private to each process and takes seconds to load, so processes that share an index or issue few queries should map the file. Furthermore, \mkc{} supports only $k\le 32$, its index is static and provides neither sorted iteration nor set operations, an in-memory counter is faster on inputs of a few megabases, and its temporary files are up to 4.4 times as large as those of \kmc{} on genomes. The fast kernels require AVX2, and the scalar path reaches about half to two thirds of their speed. Finally, all measurements come from one AMD Zen~3 workstation with a fast NVMe drive, and on slower storage the temporary I/O of any disk-based counter becomes the bottleneck.

\section{Conclusion}\label{sec:conclusion}
We have presented \mkc, an exact \kmer{} counter that builds a static two-choice cuckoo index of counts in the manner of a disk-based counter. On \NumDatasets{} genomes and read sets, \mkc{} builds its index \BuildSpeedKmcMin--\BuildSpeedKmcMax{} times faster than \kmc{} and up to \BuildSpeedHgMax{} times faster than \hg, has a smaller resident set than each of them in its faster setting on every input except the SARS-CoV-2 collection, and writes the smallest index of the four. It answers random point queries \LikeSmallHgRandomMin--\LikeSmallHgRandomMax{} times faster than the fastest hash-based alternative, and its batch interface answers queries in sequence order \BatchHgSpreadMin--\BatchHgSpreadMax{} times faster than \hg. In the future, the same offline assembly can be applied to other static \kmer{} tables, since all bits of a record beyond the quotient and the choice bit can hold taxon identifiers, colors or positions instead of counts, and filtering by a minimum count between the spool and the assembly would reduce the size of read-set indexes several-fold. \mkc{} is written in C99, runs on Linux and Windows, and comes with a header-only reader and Python bindings. Its source code is available at \url{https://github.com/iczelia/mkc} under the GNU General Public License, version~3, and the supplementary material contains the harness, setup and download scripts and raw results that reproduce every measurement.


\begin{IEEEbiographynophoto}{Kamila Szewczyk}
is a graduate student and a research assistant with the Algorithmic
Bioinformatics group at Saarland University, Saarbr\"ucken, Germany,
and the Department of Mathematics at Saarland University, Saarbr\"ucken,
Germany. Her research interests include data compression, information
theory, algorithms, compiler design, and molecular genomics.
\end{IEEEbiographynophoto}

\end{document}